\documentstyle[11pt]{article}
 \def\TH{\Theta}
\def\pa{\partial}
\def\g{\gamma} \def\G{\Gamma}
\def\a{\alpha} 
\def\b{\beta} 
\def\d{\delta} \def\D{\Delta}
\def\e{\epsilon} 
\def\z{\zeta} 

\def\l{\lambda} 
\def\m{\mu} 
\def\n{\nu} 
\def\x{\chi} 
\def\p{\pi} 
\def\v{\varphi}
 
\def\s{\sigma} 
\def\t{\tau}

\def\mn{{\mu\nu}}
\def\be{\begin{equation}}
\def\ee{\end{equation}}
\def\cA{{\cal A}}
\def\cB{{\cal B}}
\def\oU{{\bar U}}
\def\cF{{\cal F}}
\def\cG{{\cal G}}
\def\la{L_{\a}}
\def\laa{L_{\a\a}}
\def\lab{L_{\a\b}}
\def\lbb{L_{\b\b}}
\def\laaa{L_{\a\a\a}}
\def\laab{L_{\a\a\b}}
\def\labb{L_{\a\b\b}}
\def\lbbb{L_{\b\b\b}}
    
\begin{document}

\begin{titlepage}

\begin{flushright}

BRX TH-448\\

\end{flushright}

\begin{center}

{\large\bf Shock-Free Wave Propagation in Gauge Theories}

\end{center}
\vspace{3cm}

\begin{center}
{ J. McCarthy \footnote{Jim.McCarthy@dsto.defence.gov.au $\,\,\,$
  Present Address: DSTO (ITD), PO Box 1500, Salisbury SA 5108, Australia}\\
  {\sl Department of Physics and Mathematical Physics, University
       of Adelaide, SA 5005, Australia}\\
  and\\
  \"{O}. Sar{\i}o\~{g}lu \footnote{sarioglu@binah.cc.brandeis.edu}\\
  {\sl Department of Physics, Brandeis University, Waltham, MA 02454-9110, USA}}
\end{center}

\vspace{3cm}

\begin{abstract}
We present the shock-free wave propagation requirements for 
massless fields. First, we briefly argue how the ``completely
exceptional" approach, originally developed to study the 
characteristics of hyperbolic systems in $1+1$ dimensions, can be
generalized to higher dimensions and used to describe propagation
without emerging shocks, with characteristic flow remaining parallel
along the waves. We then study the resulting requirements for
scalar, vector, vector-scalar and gravity models and characterize
physically acceptable actions in each case. 
\end{abstract}

\end{titlepage}

\section{Introduction \label{intro}}
\renewcommand{\theequation}{1.\arabic{equation}}
\setcounter{equation}{0}

In this work, a brief version of which appeared in \cite{djs}, 
we study the propagation of excitations of classical
massless field actions. In
general, criteria for physical propagation of such waves can be
derived in many ways. Here, we will only consider the ``completely
exceptional" (CE) approach \cite{lax}, originally 
developed for systems in $D=1+1$. Roughly speaking, complete
exceptionality is the property ensuring that the initial
``wavefronts" evolve so as to prevent
the emergence of shocks, which, in general, result when the
``characteristics" propagate at different speeds.
As we are not aware of a rigorous procedure extending ideas 
developed at $D=2$ to higher $D$, 
we will follow steps similar to those in $D=2$, and then outline 
how to generalize them to higher
dimensions. In the process, we show how the CE idea can be looked at in 
seemingly different ways and
outline a derivation that fills the gap between the two
viewpoints. We apply our criteria to massless spin 
0,1,2 nonlinear systems. 

We start, in Section \ref{forma}, by introducing the type of physical problems that we
will study and develop the formalism that will be used throughout. 
Section \ref{anal} gives the analysis of characteristic
surfaces, which are crucial to the CE idea. In Section \ref{exam}, we
give a simple example in $D=2$, and demonstrate how shocks may
be prevented for this particular problem. Motivated by this example,
we next show how the introduced ideas can be extended to higher
dimensions in Section \ref{excep}. This naturally leads to the CE concept and
we show how one can view it in two seemingly different ways, which
are explained in the text. In Section \ref{scalar}, we study in detail the 
scalar field in $D=4$ using these two separate methods, 
derive the CE condition on it and argue as to how one can generalize 
the result to arbitrary $D$. Next, we turn to models of 
nonlinear electrodynamics in Section \ref{ned}. Here we encounter 
particular models, the constraints on which not only 
automatically guarantee the CE property (as originally discussed in \cite{boi}), 
but also ensure that both polarizations of light propagate according to the same
dispersion law, i.e. ``no birefringence" \cite{ple,bb}. Hence we call
these constraints the ``strong CE" conditions. We also derive (for the first 
time to our knowledge) the regular CE requirement conditions (much weaker than strong CE) 
in the most general $D=4$ case. Finally, in Section \ref{grav}, we find that wide 
classes of gravity models share with Einstein the null nature of 
their characteristic surfaces. In three Appendices, we show the details 
of some calculations skipped in the text.

\section{The Formalism \label{forma}}
\renewcommand{\theequation}{2.\arabic{equation}}
\setcounter{equation}{0}

In this paper we will be dealing with systems of PDEs that are Euler-Lagrange
equations of relativistic actions. 
They will be linear in highest derivatives (quasilinear) and their coefficients
will not depend on the coordinates explicitly. So, they can be reduced to a set of differential equations of first derivative 
order. Hence for $U$ an $N$-vector of fields, $\cA$ an
$N \times N$ matrix and $\cB$ an $N$-vector (both arbitrary smooth functions of
$U$), the equations of interest can always be written in the form
\be
\cA^{\m}(U) \, (\pa_{\m} U) + \cB (U) = 0 \; . \label{gen}
\ee

The theory of such equations in arbitrary dimensions is quite difficult,
but we will be mainly interested in the evolution of the spatial boundary of a wave 
propagating into some given vacuum. So, with $\oU$ some smooth (say at least
$C^1$) solution, at some initial time we have some spatial region outside of which
the ``state" is the ``vacuum solution" $\oU$, and across the boundary surface the
full solution $U$ is continuous but its first derivative may not be. We want to
consider the evolution of such initial ``wavefronts".

We will follow the formalism developed for this situation in \cite{lax,boi} and 
the references therein. Let the hypersurface, $S$, specified by  
\be
\v(x^\m)=0 \; , \label{fi=0}
\ee
denote the surface of evolution of the initial wavefront; i.e. the initial wavefront
is the spatial surface $\v(0,\vec{x})=0$. Assume that the field $U$ is continuous across
$S$; so only the normal derivative can be discontinuous. Choosing a
local coordinate system denoted by $x^{\m}=(\v,\psi^i)$, the ``first order 
discontinuity" in a given quantity $f$ can be defined as
\be
\d_1 f \equiv \left[ \frac{\pa f}{\pa \v} \right] \; , \label{d1f}
\ee
where
\be
[X] \equiv X|_{\v = 0^{+}} - X|_{\v = 0^{-}} \equiv \d_0 X \;\;. \label{[x]}
\ee
Then it is easy to check that
\be
\d_0 \, (\pa_{\m} U) = [\pa_{\m} U] = (\pa_{\m} \v) \, \d_1 U \; . \label{[dU]}
\ee

Here we are considering the possibility that $\d_1 U$ is discontinuous. Taking 
``first order discontinuity" is then like differentiation,
\be
\d_1 f(U) = (\nabla_{U} f) \, \d_1 U \; . \label{delU}
\ee
The generalization to quasilinear systems of higher order, say $q$, in derivatives
is straightforward now. Define
\be
\d_r f \equiv \left[ \frac{\pa^r f}{\pa \v^r} \right] \; , \label{drf}
\ee
and consider the case that
\be
\d_q U \neq 0 \;\; , \;\; \d_r U = 0 \;\; , \; 0 \leq r < q \; . \label{dqU}
\ee
Notice that ``taking the discontinuity" depends on the order of derivative. For
example, if $f$ has a second order discontinuity, i.e. $\d_2 f \neq 0$, then
$\d_1 f = 0$, but $\d_1 (\pa_{\m} f) \neq 0$. Hence in general one has 
\be
\d_r \, \pa_{\m} = (\pa_{\m} \v) \, \d_{r+1} \; . \label{drd}
\ee

\section{Analysis of the Characteristics \label{anal}}
\renewcommand{\theequation}{3.\arabic{equation}}
\setcounter{equation}{0}

Taking the discontinuity of (\ref{gen}), we find that, on $S$,
\be
(\cA^{\m} \v_{\m}) \, \d_1 U = 0 \; . \label{disgen}
\ee
where $\cA^{\m} = \cA^{\m} (\oU)$.
[Here $\v_{\m} \equiv \pa_{\m} \v$; henceforth, we will drop the subscript $1$
on $\d$ and use $\d U$ to mean the first order discontinuity in $U$, i.e. $\d_1 U$.]
Since $\d U \neq 0$, we see that $S$ must be a characteristic surface, i.e.
\be
{\rm det} \, (\cA^{\m} \v_{\m}) = 0  \label{detcon}
\ee
must hold on $S$. Thus $\d U$ is in the kernel of $\cA^{\m} \v_{\m}$, for a given
choice of root in (\ref{detcon}).

We can assume that (\ref{gen}) can always be rewritten such that $\cA^0$ is the
identity matrix and also that we have a flat metric on spacetime. We next define the
unit normal to $S$,
\be
{\bf \hat{n}} = \frac{\vec{\nabla} \v}{| \vec{\nabla} \v |} \; , \label{nhead}
\ee
and the ``characteristic eigenvalue"
\be
\l = - \, \frac{\pa_0 \v}{| \vec{\nabla} \v |} \; . \label{lamb}
\ee

So for a given choice of root, $\d U$ is always a linear combination of the right eigenvectors of
$\cA_n \equiv {\bf \hat{n}} \cdot \vec{\cA}$ for the corresponding eigenvalue $\l$.
In the hyperbolic case, the set of eigenvalues, $\l^{(I)} (I=1, \ldots ,N)$, are 
distinct, and the corresponding right (left) eigenvectors $R^I \, (L^I)$ are real and form a linearly independent set. 

For general $\cA^{\m}$ ($\cA^{0}$ not necessarily equal to the identity matrix),
$\l$ are just the roots of the characteristic equation (\ref{detcon}) and we have
\be
- L_I \cA^0 \l^{(I)} + L_I \cA_n = 0 = \cA_n R_J - \l^{(J)} \cA^0 R_J \;\; ; \label{ein1}
\ee
\be
- L_I \cA^0 \l^{(I)} R_J + L_I \cA_n R_J = 0 = L_I \cA_n R_J - L_I \l^{(J)} \cA^0 R_J \;\; . \label{ein2}
\ee
Then for the hyperbolic case, $L_I \cA^0 R_J = 0$, $(I \neq J)$, and one can
always choose to normalize such that

\be
L_I \cA^0 R_J = \d^{IJ} \;\; . \label{ortho}
\ee

The characteristic equation (\ref{detcon}) is homogeneous of order $N$ in $p_{\m} \equiv \v_{\m} $. By analogy, we can write it as 
\be
H(x,p)=0 \; , \label{H=0}
\ee
where we may introduce the explicit coefficients
\be
H(x,p)= G^{\m_1 \ldots \m_{N}} \, p_{\m_1} \ldots p_{\m_{N}} \;\;\; . \label{coef}
\ee
Then, by homogeneity of $H$,
\be
\sum_{\m} p_{\m} \frac{\pa H}{\pa p_{\m}} = N H = 0 \;\; , 
\ee
which can be written as
\be
\vec{p} \cdot \vec{\nabla}_{p} H = 0
\ee
where $\vec{p}$ is the $D$ dimensional vector with components
$(p_0, \ldots , p_{(D-1)})$ and $\vec{\nabla}_{p} H$ is the vector
with components $(\pa H/\pa p_0, \ldots , \pa H/\pa p_{(D-1)})$.

Since $\vec{p}$ is the normal to the hypersurface $S$, we see that the tangential
vector is parallel to $\vec{\nabla}_{p} H$, or that the curves
\be
\frac{d x^{\m}}{d s} = \frac{\pa H}{\pa p_{\m}} \label{dxs}
\ee
are tangential on (\ref{H=0}). Notice that this is a set of curves, one for each root of the characteristic equation. In analogy to classical mechanics, the ``momenta" then satisfy
\be
\frac{d p_{\m}}{d s} = - \, \frac{\pa H}{\pa x^{\m}} \label{dps}
\ee
on $S$. This follows from $\frac{d H}{d s} = 0$ for a tangential deformation and from the compatibility condition $\pa_{\m} p_{\n} - \pa_{\n} p_{\m} = 0$. We can
eliminate $s$ for $t=x^0$, and then $H$ factors as 
\be
H = \prod_{I=1}^{N} (p_0 - h^I) \; , \label{prod}
\ee
and $p_0$ can be fixed as one of the roots. Reparametrizing, for a given root $p_0 = h^{I_0} (U,p_i)$, we can span the characteristic surface with the trajectories obeying
\be
\frac{d x^{\m}}{d s} = \frac{\pa h^{I_0}}{\pa p_{\m}} \;\;\;\; , \;\;\;\;
\frac{d p_{\m}}{d s} = - \, \frac{\pa h^{I_0}}{\pa x^{\m}} \;\; . \label{xps}
\ee

\section{An Example \label{exam}}
\renewcommand{\theequation}{4.\arabic{equation}}
\setcounter{equation}{0}

Consider the following $D=2$ example \cite{lax}.
Take the simple PDE
\be
\pa_t u + u \, \pa_x u = 0 \; . \label{udenk}
\ee
Then it easily follows that the characteristic curve is
\be
\frac{d x}{d t} = u(x,t) \; . \label{xtu}
\ee
Clearly (\ref{udenk}) and (\ref{xtu}) imply $\frac{d u}{d t}=0$ along the characteristic curve, i.e. the characteristic curve is a line with constant $u$. So the ``velocity",
$\frac{d x}{d t}$, is constant and the characteristics are nothing but straight lines. If 
we denote by $\v$ the point where the given line is at initial time $t=0$, then by integrating (\ref{xtu}),   
\be
x(t) = u(\v,0) \, t + \v \; . \label{xsol}
\ee
This implicit equation for $\v = \v(x,t)$ is just the equation for the given characteristic curve, parametrized by its initial point. So then one has
\be
u(x,t) = u(\v(x,t),0) \;\; , \label{usol}
\ee
for the solution to (\ref{udenk}), in terms of the initial value of $u$ at $t=0$.

For a linear PDE, the coefficient of $\pa_x u$ in (\ref{udenk}) is a constant independent of $u$ and the characteristic curves are parallel straight lines. In the general case, when the coefficient of $\pa_x u$ in (\ref{udenk}) is an arbitrary, say smooth, function of $u$, the
slope of a given characteristic curve depends on the initial value of $u$ at the starting  point. Thus as time evolves, the characteristic curves can intersect and a shock may develop. It seems that this can be prevented only if the ``velocity", $\frac{d x}{d t}$, can be  made independent of the coordinate normal to the characteristic.

\section{Exceptionality \label{excep}}
\renewcommand{\theequation}{5.\arabic{equation}}
\setcounter{equation}{0}

Even though we have neither found a proof in the literature nor been able to
prove it rigorously, this ``method of characteristics" seems to extend to first order PDEs in higher dimensions . Although there doesn't seem to be such a construction for the matrix system (\ref{gen}), we want to carry on our discussion and see what can be done.

Motivated by this 2-dimensional example, let us look for situations where the
characteristic surfaces do not cross as they evolve, hence shock waves do not
develop. Following the reasoning given above, we can demand that (locally) the
characteristic eigenvalue, which after all is analogous to the ``velocity" in
the given example, is independent of $\v$ in the evolution, or that 
\be
\frac{\pa \l}{\pa \v} = 0 = (\nabla_U \l) \, \frac{\pa U}{\pa \v} \;\; . \label{dlaph}
\ee

Now let us look at the homogeneous case in (\ref{gen}) (i.e. $\cB=0$). Then taking a  particular root $p_0=h^{(I_0)}$ of the characteristic equation (\ref{prod}), defines a family of surfaces (remember $p_{\m} \equiv \pa_{\m} \v^{(I_0)}$)
\be
\v^{(I_0)} (0,\vec{x}) =  {\rm constant} \; . \label{vcon}
\ee
Assume that $U$ is just a function of $\v$. Then (\ref{gen}) implies (by $\cB=0$) that
\be
(\cA^{\m} \v_{\m}) \, \frac{d U}{d \v} = 0 \; , \label{uphi}
\ee
and hence
\be
\frac{d U}{d \v} = \x^{(J)} (\v) \, R^{(J)}   \, ,  \label{chiR}
\ee
where $R^{(J)}$ is the corresponding right eigenvector for the given root. 
$U$ and $R^{(J)}$ being $N$-vectors, we end up with $N$ ordinary differential equations. 
We can always assume that $R^{(J)}$ has a nonzero component, and that particular component $U_K$ can always be chosen such that it is equal to $\v$. Since the eigenvector $R^{(J)}$ is known as a function of $U$ and $\v$, the ratios of the other components determine $U_A = 
U_A (U_K) \, (A \neq K)$. The particular component $U_K$ itself can be determined from the characteristic equation. Now a solution for $U$ obtained in this way is called a {\it simple wave} \cite{lax}.

This brings us to the so called exceptionality condition \cite{lax}. Let us
first define what is meant by that:

The wave corresponding to a given characteristic root is called {\it exceptional} if it is
such that
\be
(\nabla_U \l) \, \cdot \, R = 0 \;\; . \label{norm}
\ee

Moreover when all the $N$ wave modes are exceptional, the system is said to be CE \cite{lax}.

So for simple wave solutions of the homogeneous case just discussed, exceptionality
condition is just the statement that $\nabla_U \l$ is orthogonal to the corresponding
right eigenvector. In light of (\ref{dlaph}) and the discussion that led to it,
exceptionality condition does after all seem to ``justify" a generalization of the
(naive) idea we developed to prevent the development of shocks using the $D=2$ example
(at least for the case of simple waves).

Another way of looking at the problem may be provided by the following:

From (\ref{disgen}), it follows that $\d U$ can be expressed as a linear combination of 
right eigenvectors as 
\be
\d U = \p^I \, R_I \label{piR}
\ee
for some components $\p^I \, (I=1, \ldots, N)$ (also called as the coefficients of discontinuity). In general, one would expect these coefficients to evolve according to a nonlinear differential equation. In Appendix A, it is shown, how the CE condition
can also be viewed as the statement that the coefficients of discontinuity evolve
according to a linear ODE, thus the characteristic curves are prevented from intersecting
locally.

We now briefly mention another alternative approach developed in \cite{boi}
for a ``covariant formulation" of exceptionality. Let us choose a particular
root $p_0=h^{(J)} (U,p_i)$ for some $J$ in (\ref{prod}). We also have by
(\ref{lamb}) that $p_0$ is proportional to $\l$. If we now take 
the field-gradient, $\nabla_U$, of the ``Hamiltonian" $H$ in (\ref{prod}) and
set $p_0=h^{(J)}$ afterwards, we see that only the term which is proportional 
to $\nabla_U p_0 \sim \nabla_U \l$ does not vanish in the resultant expression.
For a simple wave then, contracting this with $\d U$ and using (\ref{delU}),
we get
\be
\nabla_U \l \cdot \d U = \d \l   \label{dlamb}
\ee
for this particular root.

Hence, in light of (\ref{dlaph}) and (\ref{norm}), one arrives at the 
``equivalent" condition for exceptionality:

The wave corresponding to a given root is exceptional, if on the characteristic surface $H=0$, one has
\be
\d \l = 0 \;\; . \label{dlam}
\ee

Again for CE, this must hold for all roots, or that
\be
\d H = 0 \;\; . \label{dH=0}
\ee

In the following, we apply the above mentioned (two seemingly different) 
CE requirements in a variety of physical cases. In the process we supply 
the missing details leading to the results reported in \cite{djs}. We want to
make it clear that (\ref{norm}) was originally developed for systems in $D=2$ only
\cite{lax}. Here, however, we apply (\ref{norm}) and (\ref{dH=0}) to systems in higher
dimensions. Although we are not aware of a rigorous construction that generalizes the 
results explained so far to PDEs in higher $D$, it is plausible that
such a general proof can be given.

After all, notice that the characteristic 
equation, and the condition for CE, are algebraic equations
which must hold pointwise in any $x^{\mu}$. At a fixed point
on the characteristic surface at a fixed time, the normal
${\bf \hat{n}}$ is a fixed vector and proceeding for arbitrary
${\bf \hat{n}}$, and $U$, is the same as imposing the conditions
pointwise. Furthermore, the original system is rotationally invariant,
where rotations act on $U$ as some linear matrix
representation. So, the CE conditions are rotation invariant,
and having chosen ${\bf \hat{n}}$ (i.e. working at a fixed point and
time) we can just rotate it to, say, the first coordinate
direction $x^{1}$ and proceed to study the eigensystem
$|{\bf \cA}^{1} - \l \, {\bf I}| = 0$, provided the system
can be brought into a form such that ${\bf \cA}^0$ equals the identity matrix. 
Of course $U$ changes in rotating, but the eigensystem is worked out 
for arbitrary $U$.

\section{Scalar Field \label{scalar}}
\renewcommand{\theequation}{6.\arabic{equation}}
\setcounter{equation}{0}

We now want to study in detail the CE requirement for a scalar field in $D=4$. 
We first work out the problem using the requirement (\ref{norm}), then show that one
finds the same answer (with considerably less effort) using condition (\ref{dH=0}),
as was in fact done earlier in \cite{boi}.

\subsection{The First Way}

Given the covariant action \( I = \int d^4 x \, L(z) \), 
where $z \equiv \frac{1}{2} (\pa_{\m} \s)^2$ is the
only invariant (in first derivatives), $\eta=(-,+,+,+)$, the 
field equations can be written as 
\be
\pa_{\m} \left( (\pa^{\m} \s) L^\prime \right) =
(\pa_{\m} \pa^{\n} \s) \, (\pa_{\n} \s) \, (\pa^{\m} \s) \, L^{\prime\prime}
+ (\pa_{\m} \pa^{\m} \s) \, L^\prime = 0 \;\; . \label{scafeq} 
\ee
Here $^\prime$ denotes differentiation with respect to $z$.

By defining $A \equiv \pa_{0} \s$, $B \equiv \pa_{1} \s$,
$C \equiv \pa_{2} \s$ and $D \equiv \pa_{3} \s$ (hence
$z= \frac{1}{2} (-A^2+B^2+C^2+D^2)$), we can take  
${\bf U} = (A,B,C,D)$ and write this system in canonical form as
\[ {\bf I} \; \frac{\pa {\bf U}}{\pa t} + 
{\bf M}^{i} \; \frac{\pa {\bf U}}{\pa x^{i}} = {\bf 0} \]
where each ${\bf M}^{i}$ has elements (with $i=1,2,3;\m=0,1,2,3$) 
\[ m^{1}_{00}=\frac{-2 A B L^{\prime\prime}}{\TH} \;,
   m^{1}_{01}=\frac{B^2 L^{\prime\prime} + L^\prime}{\TH} \;,
   m^{1}_{02}=\frac{B C L^{\prime\prime}}{\TH} \;,
   m^{1}_{03}=\frac{B D L^{\prime\prime}}{\TH} \;; \]
\[ m^{1}_{2\m}=m^{1}_{3\m}=0 \;; m^{1}_{1\m}=-\d_{0\m} \;; \]
\[ m^{2}_{00}=\frac{-2 A C L^{\prime\prime}}{\TH} \;,
   m^{2}_{01}=m^{1}_{02} \;,
   m^{2}_{02}=\frac{C^2 L^{\prime\prime} + L^\prime}{\TH} \;,
   m^{2}_{03}=\frac{C D L^{\prime\prime}}{\TH} \;; \]
\[ m^{2}_{1\m}=m^{2}_{3\m}=0 \;; m^{2}_{2\m}=-\d_{0\m} \;; \]
\[ m^{3}_{00}=\frac{-2 A D L^{\prime\prime}}{\TH} \;,
   m^{3}_{01}=m^{1}_{03} \;,
   m^{3}_{02}=m^{2}_{03} \;,
   m^{3}_{03}=\frac{D^2 L^{\prime\prime} + L^\prime}{\TH} \;; \]
\[ m^{3}_{1\m}=m^{3}_{2\m}=0 \;; m^{3}_{3\m}=-\d_{0\m} \]
and $\TH \equiv A^2 L^{\prime\prime} - L^\prime$. Here we have also used the compatibility conditions \( \frac{\pa B}{\pa t} = \frac{\pa A}{\pa x^{1}} \; ,
\frac{\pa C}{\pa t} = \frac{\pa A}{\pa x^{2}} \; \) 
and \( \frac{\pa D}{\pa t} = \frac{\pa A}{\pa x^{3}} \).

So, by the reasoning given at the end of the last section, we proceed to
impose the CE condition (\ref{norm}) using the eigensystem 
$|{\bf M}^{1} - \l \, {\bf I}| = 0$. \footnote{In fact, we showed separately that
taking arbitrary ${\bf \hat{n}}$ does not alter the final results obtained in this
section.} The characteristic polynomial of ${\bf M}^{1}$ turns out
to be $\l^2 (\l^2 + a_1 \l + a_2) = 0$ where
$a_1 \equiv \frac{2 A B L^{\prime\prime}}{\TH}$ and
$a_2 \equiv \frac{B^2 L^{\prime\prime} + L^{\prime}}{\TH}$. Apart from the 
eigenvalue at $\l =0$ (with multiplicity 2), there are
two distinct eigenvalues $\l_3,\l_4$ in the general case.
\footnote{For the degenerate case $L^{\prime} [L^{\prime} - (A^2-B^2) L^{\prime\prime}]=0$,
$\l_3=\l_4=-\frac{A}{B}$ but then there is no nontrivial covariant 
action which can satisfy this. Moreover in this case, the 
system is no longer hyperbolic.} The eigenvectors corresponding 
to each can be taken as
\[ {\bf e}_1 = \left( 0,\frac{-B C L^{\prime\prime}}{B^2 L^{\prime\prime} + L^{\prime}},1,0 \right)^{T} \;\; , \;\;
   {\bf e}_3 = \left( -\l_3,1,0,0 \right)^{T} \;, \]
\[ {\bf e}_2 = \left( 0,\frac{-B D L^{\prime\prime}}{B^2 L^{\prime\prime} + L^{\prime}},0,1 \right)^{T} \;\; , \;\;
   {\bf e}_4 = \left( -\l_4,1,0,0 \right)^{T} \]
which clearly form a full linearly independent set, hence
our system is hyperbolic. We next apply the CE condition (\ref{norm})
to this eigensystem. Obviously,
it will be trivially satisfied for $\l = 0$. For the
remaining nontrivial eigenvalues, note that by differentiating
$\l^2 + a_1 \l + a_2 = 0$, we can write
\[ \frac{\pa \l}{\pa U_s} = 
- \, \frac{\l \frac{\pa a_1}{\pa U_s} + \frac{\pa a_2}{\pa U_s}}
{2 \l + a_1} \]
and the CE condition
\( \sum_{s} \frac{\pa \l_{p}}{\pa U_s} \; e_{p, \, s} = 0 \)
becomes
\[ \l^2 \frac{\pa a_1}{\pa A} + \l ( \frac{\pa a_2}{\pa A} -
\frac{\pa a_1}{\pa B} ) - \frac{\pa a_2}{\pa B} = 0 \;\;\;\;
{\rm for} \;\; \l=\l_3 , \l_4 \]
by using the explicit form of the eigenvectors. However we 
know that $\l_3 , \l_4$ satisfy $\l^2 + a_1 \l + a_2 = 0$.
Hence these two equations must be linearly dependent
which implies that
\[ a_1 \frac{\pa a_1}{\pa A} + \frac{\pa a_1}{\pa B} -
\frac{\pa a_2}{\pa A} = 0 \;\;\; {\rm and} \;\;\;
a_2 \frac{\pa a_1}{\pa A} + \frac{\pa a_2}{\pa B} = 0 \]
have to be satisfied simultaneously.

Substituting the explicit forms of $a_1$ and $a_2$,
we find after some calculation that
\begin{eqnarray} 
a_1 \frac{\pa a_1}{\pa A} + \frac{\pa a_1}{\pa B} -
\frac{\pa a_2}{\pa A} & = & \frac{L^{\prime} L^{\prime\prime\prime} - 3 (L^{\prime\prime})^2}{\TH^3} \,
[ A (3 B^2 + A^2) L^{\prime} + A^3 (B^2 - A^2) L^{\prime\prime} ] = 0 \; , \label{con1} \\
a_2 \frac{\pa a_1}{\pa A} + \frac{\pa a_2}{\pa B} & = &
\frac{L^{\prime} L^{\prime\prime\prime} - 3 (L^{\prime\prime})^2}{\TH^3} \; [ B (3 A^2 + B^2) L^{\prime} + 
B A^2 (B^2 - A^2) L^{\prime\prime} ] = 0 \; . \label{con2}
\end{eqnarray} 

The only nontrivial covariant condition we can impose
such that these two constraints are satisfied simultaneously is
\be
L^{\prime} L^{\prime\prime\prime} - 3 (L^{\prime\prime})^2 = 0  \; . \label{scal}
\ee

\subsection{The Second Way}

In this part, we want to impose (\ref{dH=0}) using the formalism developed starting in Section \ref{forma}. We now look for a surface $S$ across which the discontinuity in $\s$ is second order. Thus with $\d_2 \s \equiv Q$, we have
($\s_{\m} \equiv \pa_{\m} \s \, , \v_{\m} \equiv \pa_{\m} \v$)
\[ \d \s_{\m} = \v_{\m} \, Q \;\; . \]
Taking the discontinuity of (\ref{scafeq}) gives
\be
\v_{\m} \left( (\d L^\prime) \, \s^{\m} + (\d \s^{\m}) \, L^\prime \right) = 0 \label{kil1}
\ee
which, with $\d z = \s^{\m} \, (\d \s_{\m}) = \s^{\m} \, \v_{\m} \, Q$ and 
$\d L^\prime = L^{\prime\prime} \, \d z$, becomes
\be
Q \, \left( \cG \, L^\prime + (\s^{\n} \v_{\n})^2 L^{\prime\prime} \right) = 0  \label{kil2}
\ee
where $\cG \equiv \v^{\m} \v_{\m}$. Comparing this to the previous discussion, we have $H(x,p) = G^{\mn} p_{\m} p_{\n} = 0$ with $(p_{\m}=\v_{\m} , Q \neq 0)$
\be
G^{\mn} = \eta^{\mn} \, L^\prime + \s^{\m} \, \s^{\n} \, L^{\prime\prime} \;\; . \label{kil3}
\ee

Imposing (\ref{dH=0}) (taking the discontinuity) gives
\be
Q \, \left( 3 \, \cG \, L^{\prime\prime} + (\s^{\l} \v_{\l})^2 \, L^{\prime\prime\prime} \right) \, (\s^{\m} \v_{\m}) = 0  \label{kil4}
\ee
and using $\cG \, L^\prime + (\s^{\n} \v_{\n})^2 L^{\prime\prime} = 0$ in (\ref{kil4}) yields
\be
Q \, \cG \, \left( - \, \frac{L^\prime \, L^{\prime\prime\prime}}{L^{\prime\prime}} + 3 \, L^{\prime\prime} \right) \, (\s^{\n} \v_{\n}) = 0 \label{kil5}
\ee
This again leaves us with the condition (\ref{scal}).

Notice that throughout, we have never used the fact that $D=4$. This suggests that (\ref{scal}) is a $D$-invariant $(D \geq 2)$ condition. For general $D$, using
the requirement (\ref{norm}), one ends up with ${\bf M}^{i}$ which individually have 
$\l = 0$ (with multiplicities $D-2$) and the remaining two nontrivial eigenvalues (corresponding to the pair of canonical variables $(\s, \pi)$ for the only degree of freedom of the theory) with their corresponding eigenvectors yield (\ref{scal}) when inserted into (\ref{norm}).

To find the solutions of (\ref{scal}), we first note that by 
defining $X \equiv L^\prime$, we can write it as 
\( X^4 \left( \frac{X^\prime}{X^3} \right)^\prime = 0 \) which will be
satisfied nontrivially provided $X^\prime=0$ or 
\( \left( \frac{X^\prime}{X^3} \right)^\prime = 0 \; . \) Integrating
these simple equations, we find $X=c_1 \;, L=c_1 z + c_2$ or
\( \left( \frac{1}{X^2} \right)^\prime = -2 c_3 \; , 
\frac{1}{X^2} = -2 c_3 z + c_4 \; , 
L= \pm \frac{1}{c_3} \sqrt{-2 c_3 z + c_4} + c_5 \) for
$c_{q} (q=1, \dots ,5)$ arbitrary integration constants. Choosing
these constants suitably, we note the particularly interesting
cases as $L=-z=-\frac{1}{2} (\pa_{\m} \s)^2$ and
\( L = 1-\sqrt{1+2z} = 1-\sqrt{1+ (\pa_{\m} \s)^2} =
1-\sqrt{-\det[\eta_{\m\n}+(\pa_{\m} \s) \, (\pa_{\n} \s)]} \), 
the scalar analogs to Maxwell and Born-Infeld electrodynamics, 
respectively.

\section{Nonlinear Electrodynamics in $D=4$ \label{ned}}
\renewcommand{\theequation}{7.\arabic{equation}}
\setcounter{equation}{0}

We now come to, our most physically important example, the $D=4$ 
Abelian gauge vector theories. Any gauge invariant action, depending on
$F_{\m\n}= \pa_{\m} A_{\n} - \pa_{\n} A_{\m}$ but not its derivatives, has the form
\be
I[A_{\m}] = \int d^4 x \; L(\a,\b) \;\;, \;\;
\a \equiv \frac{1}{2} F_{\m\n} F^{\m\n} \;\;, \;\;
\b \equiv \frac{1}{4} F_{\m\n} ~^*\!F^{\m\n} \;\;, \;\;
^*\!F^{\m\n} \equiv \frac{1}{2} \e^{\m\n\s\t} F_{\s\t} \; . \label{act1}
\ee
Here subscripts on $L$ mean differentiation with respect to the
(only possible) invariants $\a$ or $\b$ 
and with our conventions $\e^{0123}=+1$,
$\eta_{\mu\nu} =(-,+,+,+)$, $\a={\bf B}^2 - {\bf E}^2$, 
$\b=-{\bf B} \cdot {\bf E}$ with $E^{i} \equiv F^{0i}$ and
$B^{i} \equiv \frac{1}{2} \e^{ijk} F_{jk}$.

We first drop the $\b$-dependence of $L$, show in detail how the CE
condition (\ref{dH=0}) is applied to $L(\a)$, then reinclude $\b$ and carry out the
CE condition (\ref{dH=0}) for full $L(\a,\b)$. [Again we originally studied this problem using the requirement (\ref{norm}) which is quite laborious and tedious. We show in the Appendix B, the general outline of how (\ref{norm}) is carried out for $L(\a)$. We don't show how (\ref{norm}) is applied to the most general case, $L(\a,\b)$, although in this case we were able to prove, at least, the sufficiency of (\ref{vec1}) and (\ref{vec2}) using (\ref{norm}).]

We look for a hypersurface $S$ across which the discontinuity in $A_{\m}$ is second order. Hence, with $\d_2 A_{\m} = \p_{\m}$, we have
\be
\d F^{\m\n} = \v^{\m} \, \p^{\n} - \v^{\n} \, \p^{\m} \;\;\; , \;\;
\d \, ^*\!F^{\m\n} = \e^{\m\n\s\t} \, \v_{\s} \, \p_{\t} \label{dal1}
\ee
and
\be
\d \a = 2 \, F^{\m\n} \, \v_{\m} \, \p_{\n} \;\;\; , \;\;
\d \b = \, ^*\!F^{\m\n} \, \v_{\m} \, \p_{\n} \;\; . \label{dal2}
\ee

\subsection{$L(\a)$ Case}

For $L=L(\a)$ only, the field equation is simply 
\( \pa_{\n} (F^{\m\n} L^\prime)=0 \) with the Bianchi identity $\pa_{\n} \, ^*\!F^{\m\n}=0$. [Here, $^\prime$ denotes differentiation with respect to $\a$, of course.]

Taking the discontinuity of the field equation, we find
\be
- \, 2 \, U^{\m} \, \cF \, L^{\prime\prime} + (\v_{\n} \, \p^{\n}) \, \v^{\m} \, L^\prime - \cG \, \p^{\m} \, L^\prime = 0 \label{dal3}
\ee
where we have used $U^{\m} \equiv F^{\l\m} \, \v_{\l} \, , \cG \equiv \v^{\m} \v_{\m}$ and $\cF \equiv F^{\l\s} \, \v_{\l} \, \p_{\s} = U^{\s} \, \p_{\s}$. [Taking the discontinuity of the Bianchi identity, one can see that it follows automatically.] 

Now contracting (\ref{dal3}) by $-U_{\m}$ (and assuming $\cF \neq 0$ for the general case), we find
\be
H = 2 \, u \, L^{\prime\prime} + \cG \, L^\prime = 0 \label{dal4}
\ee
where we have defined
$u \equiv U^{\m} U_{\m}$. Now $\d u = 2 \, U^{\m} \, \d U_{\m} = 2 \, \cG \, \cF$
and $\d \a = 2 \, \cF$. Hence imposing (\ref{dH=0}) gives
\be
\d H = \cF \, (4 \, u \, L^{\prime\prime\prime} + 6 \, \cG \, L^{\prime\prime}) = 0
\label{dal5}
\ee
and substituting for $u$ using (\ref{dal4}), we end up with
\be
\d H = 2 \, \cF \, \cG \, \left( 3 \, L^{\prime\prime} - \frac{L^\prime}{L^{\prime\prime}} \, L^{\prime\prime\prime} \right) = 0
\label{dal6}
\ee
Hence we again find (\ref{scal}) in a new disguise, whose solutions we can immediately copy as $L (\a) = k + (d + c \a )^{\frac{1}{2}}$ (for arbitrary constants $k,d,c$) apart from Maxwell, $L=-\, \frac{1}{2} \a$ (or $L^\prime=$ constant).

We remark that in $D=3$, where $\a$ is the only invariant,
this is also the CE result, there one also has
\( \sqrt{1+\a}= \sqrt{-\det[\eta_\mn+F_\mn]} \). In $D=2$ there is of course no
propagation for any $L(\a)$ and correspondingly no 
restrictions are imposed.

\subsection{$L(\a,\b)$ Case}

We now want to study the full Lagrangian $L(\a,\b)$. For this case, the field equation is
\( \pa_{\n} \, (L_{\a} F^{\m\n} + \frac{1}{2} L_{\b}~^*\!F^{\m\n})=0 \). Taking  
the discontinuity, we find
\be
F^{\m\n} \, \v_{\n} \, (\d L_{\a}) + \v_{\n} \, (\d F^{\m\n}) \, L_{\a} +
\frac{1}{2} \, ^*\!F^{\m\n} \, \v_{\n} \, (\d L_{\b}) = 0 \;\; . \label{got1}
\ee
Using (\ref{dal1}) and (\ref{dal2}), this becomes 
\be
-U^{\m} \, (2 \, \cF \, L_{\a\a} + \x \, L_{\a\b}) + \v_{\n} \, (\v^{\m} \, \p^{\n} - \v^{\n} \, \p^{\m}) \, L_{\a} - \frac{1}{2} \, V^{\m} \, (2 \, \cF \, L_{\a\b} + \x \, L_{\b\b}) = 0 \label{got2}
\ee
where we have used $V^{\m} \equiv \, ^*\!F^{\l\m} \, \v_{\l}$ and $\x \equiv \, ^*\!F^{\l\s} \, \v_{\l} \, \p_{\s} = V^{\s} \, \p_{\s}$.

Now contracting (\ref{got2}) by $-U_{\m}$, and then by $-V_{\m}$, we get respectively:
\begin{eqnarray}
\cF \, (2 \, u \, L_{\a\a} + \cG \, L_{\a} + \b \, \cG \, L_{\a\b}) + 
\x \, \left( u \, L_{\a\b} + \frac{1}{2} \, \b \, \cG \, L_{\b\b} \right) & = & 0 \; , \label{got3} \\
\cF \, \left( 2 \, \b \, \cG \, L_{\a\a} + L_{\a\b} \, (u - \a \, \cG) \right) +
\x \, \left( \b \, \cG \, L_{\a\b} + \cG \, L_{\a} + \frac{1}{2} \, L_{\b\b} \, (u - \a \, \cG) \right) & = & 0 \; , \label{got4}
\end{eqnarray}
where we have made use of the identities $U^{\m} V_{\m} = \b \, \cG$ and $V^{\m} V_{\m} = u - \a \, \cG$.

For this system to have nontrivial $\cF$ and $\x$, the determinant of the $2 \times 2$ matrix, that comes from writing (\ref{got3}) and (\ref{got4}) as
\( \left( \cF \;\; \x \right) \, {\bf M} = {\bf 0} \), must vanish. Hence we have 
$\left( K \equiv L_{\a\a} L_{\b\b} - L_{\a\b}^2 \right)$
\be
H=u^2 K + u \cG \left[ 2 L_{\a} \left( L_{\a\a} + \frac{1}{4} L_{\b\b} \right) - \a K \right] + \cG^2 \left[ L_{\a} \left( L_{\a} + 2 \b L_{\a\b} - \frac{1}{2} \a L_{\b\b} \right) - \b^2 K \right] = 0 \; . \label{got5}
\ee
Notice that for the discriminant, one gets
\be
\frac{\D}{\cG^2} = \frac{1}{4} \left[ - L_{\a} \, (4 \, L_{\a\a} - L_{\b\b}) + 2 \, \a \, K \right]^2 + 4 \left[ - L_{\a} \, L_{\a\b} + \b K \right]^2 \;\; . \label{got6}
\ee

For the case $\D=0$, i.e. when
\begin{eqnarray}
- L_{\a} \, (4 \, L_{\a\a} - L_{\b\b}) \, + \, 2 \, \a \,
[L_{\a\a} \, L_{\b\b} - L_{\a\b}^2] & = & 0 \;\; ,  \label{vec1} \\
- L_{\a} \, L_{\a\b} \, + \, \b \, 
[L_{\a\a} \, L_{\b\b} - L_{\a\b}^2] & = & 0 \;\; ,  \label{vec2}
\end{eqnarray}
$H$ takes the form $H=K \,(u-h)^2=0$ and for $K \neq 0$, it follows that (\ref{dH=0}) is satisfied automatically. Hence any $L$ that fulfills (\ref{vec1})
and (\ref{vec2}) is CE.

The differential constraints (\ref{vec1}) and (\ref{vec2}) were actually 
found a long time ago in different contexts \cite{ple,bb,boi}. Bialynicki-Birula 
\cite{bb} discovered these equations by studying the propagation of 
weak electromagnetic waves on a strong, constant field background. 
He showed that they were necessary for both polarizations of light to
propagate according to the same dispersion law; he calls these as the ``no 
birefringence" conditions. Pleba\'{n}ski \cite{ple}, studied 
the theory of small perturbations and their discontinuities in nonlinear 
electrodynamics and considering all possible cases for the form of the 
background field (e.g. null, algebraically general) and constraining the 
system with physical conditions such as causality along the way, proved the 
necessity and sufficiency of these differential constraints for the excitations 
of light to propagate according to a single characteristic equation, with 
coinciding characteristic surfaces. Boillat \cite{boi} found these conditions 
using equation (\ref{dH=0}) explained in this paper, and demanding that it be 
expressible as a complete square as explained in (\ref{got5}) - (\ref{vec2}).

The work of Pleba\'{n}ski involves an extensive study of characteristic 
surfaces, which is what CE formulation is all about, in nonlinear electrodynamics, 
so it is not surprising that he finds (\ref{vec1}) and (\ref{vec2}) as the 
conditions to have coinciding characteristic surfaces; after all that is also what 
Boillat gets using the CE viewpoint. Bialynicki-Birula effectively allows the 
discontinuities in terms of weak disturbances about a generic background. 
It is not surprising to see that having the same dispersion
law for both polarizations implies having a single characteristic surface for the
evolution of discontinuities. Apart from these historical details, we will call 
the two conditions, (\ref{vec1}) and (\ref{vec2}), the ``strong CE" conditions from 
now on, because of this extra physical constraint that they impose on the system.

The solutions of (\ref{vec1}) and (\ref{vec2}) are important to define physically
acceptable models of electrodynamics. It is clear that the Maxwell action,
$I_{Max} = - \frac{1}{2} \int d^4 x \a$, is indeed a solution, and it was realized
in \cite{ple,bb,boi} that another is the (once again popular) Born-Infeld action
\cite{bi},
\be
I_{BI} = \int d^4 x \; (1-\sqrt{-\det[\eta_\mn+F_\mn]}) = \int d^4 x \; (1-\sqrt{1+\a-\b^2}) \; .
\label{boin}
\ee
However, these are not the only solutions, unless one further requires that they reduce
to $I_{Max}$ for weak fields. Otherwise there are additional solutions 
such as $L=\a/\b$. [As shown in \cite{djs}, without requiring the weak field condition, imposing strong CE with duality invariance (a property shared by both of these theories), singles out Maxwell and Born-Infeld.] 

Now we continue with the general case $K \neq 0$, $\D \neq 0$. For convenience, we define 
\[ P \equiv 2 \, L_{\a} \, \left( L_{\a\a} + \frac{1}{4} L_{\b\b} \right) - \a \, K  \;\;\; , \;\;
R \equiv L_{\a} \, \left( L_{\a} + 2 \, \b \, L_{\a\b} - \frac{1}{2} \, \a \, L_{\b\b} \right) - \b^2 \, K  \;\;\; , \]
\[ p \equiv 2 \, L_{\a\a} \;\; , \; q \equiv L_{\a} + \b \, L_{\a\b} \;\; , \;
r \equiv L_{\a\b} \;\; , \; s \equiv \frac{1}{2} \, \b \, L_{\b\b} \]
and rewrite $H$ as 
\be
H = u^2 \, K + u \, \cG \, P + \cG^2 \, R = 0 \label{tar1}
\ee
Now imposing (\ref{dH=0}), we find
\be
\d H = u^2 (2 \cF K_{\a} + \x K_{\b}) + u \cG (4 \cF K + 2 \cF P_{\a} + \x P_{\b})
+ \cG^2 (2 \cF P + 2 \cF R_{\a} + \x R_{\b}) = 0 \label{tar2}
\ee
where we have used $\d u = 2 \cG \cF \; , \d \a = 2 \cF$ and $\d \b = \x$.

Now using $u^2 = - \, \frac{\cG}{K} (u P + \cG R)$ (from (\ref{tar1})) and 
$\x = - \, \frac{p u + \cG q}{r u + \cG s}$ (from (\ref{got3})), we find that
(\ref{tar2}) is simplified into a form $\d H = u \cG \z_1 + \cG^2 \z_2 = 0$. Since
$K \neq 0$ and $\D \neq 0$, this implies that $\z_1$ and $\z_2$ must vanish simultaneously. [This can also be seen as the requirement that (\ref{tar1}) and
(\ref{tar2}) be linearly independent.] Finally one finds that, CE requirements (corresponding to (\ref{dH=0})) are
\begin{eqnarray}
2 K_{\a} (r P^2 - s P K - r R K) + K_{\b} (q P K - p P^2 + p R K) +
2 P_{\a} (s K^2 - r P K) \nonumber \\ 
+ K P_{\b} (p P - q K) + 2 R_{\a} (r K^2) - R_{\b} (p K^2) - 2 r P K^2 + 4 s K^3 & = & 0 
\label{tar3} \\
2 K_{\a} (r P R - s R K) + K_{\b} (q R K - p R P) - 2 P_{\a} (r R K) +
P_{\b} (p K R)  \nonumber \\
+ 2 R_{\a} (s K^2) - R_{\b} (q K^2) - 4 r R K^2 +  2 s P K^2 & = & 0 
\label{tar4}
\end{eqnarray}

In Appendix C, we give these equations in terms of $L$ and its derivatives only.
Notice that these equations are quasilinear (linear in the third order derivatives of $L$) just like (\ref{scal}). Being of third order, they are of course weaker than
(\ref{vec1}) and (\ref{vec2}). Born-Infeld, of course, satisfies these equations but
we have neither been able to solve them in the general case, nor for the more restricted situation when one also demands duality invariance. For the latter, one would expect to get two (or, with a bit of luck, only one) ordinary differential equations involving only $\a$-derivatives when one substitutes for $\b$-derivatives by using the duality invariance constraint (see \cite{dua}) and its $(\a,\b)$ derivatives, recursively. 

An application of CE, rather than strong CE, comes from
theories involving the (neutral) scalar plus the Abelian vector field, 
where possible invariants are 
\( ( \a, \b, z \; (\equiv \frac{1}{2} (\partial_{\mu} \s)^2), y \equiv \frac{1}{2} (F_\mn \s^\n)^2 ) \). For a Lagrangian $L(\a, \b, z)$, the CE conditions further require
$L_{z\a} = 0 = L_{z\b}$, which reduce it to the noninteracting 
$L(\a ,\b ) + L(z)$ form. Having the ``fully Born-Infeld" form $\sqrt{-\det [\eta_\mn + F_\mn + \s_\m \s_\n ]}$ in mind, one can consider $L(\a, \b, y, z)$. It turns out, however, 
that there are no CE actions with nontrivial dependence on the other possible variable $y \equiv \frac{1}{2} (F_\mn \s^\n)^2$. 
Thus, CE alone separates the two systems and imposes the previously stated constraints
on their forms.

\section{Gravitational Models \label{grav}}
\renewcommand{\theequation}{8.\arabic{equation}}
\setcounter{equation}{0}

Finally, we turn to gravitation. For Einstein's gravity in vacuum, 
as well as the linearized theory, the gravitational waves are CE, 
the characteristic surfaces describing discontinuities being null
(see e.g. \cite{lic}). It can be shown that this result holds for any $D>4$. 
[For $D=3$, there is of course no propagation and no restrictions
are imposed.] One can further look at pure gravitational actions 
of the form $\int d^4 x (p R^2_\mn - q R^2) \sqrt{-g}$ in $D=4$ and 
$\int d^D x f(R) \sqrt{-g}$ in $D>3$ and 
show that the same conclusion remains unchanged.

To reduce these theories to a first order system would be
inconvenient, but is fortunately made unnecessary by a simple
extension of the previous discussion. Clearly, if we rebuilt
the original higher order equations from the set (\ref{gen}),
we would simply have the situation that all the derivatives
of the field are assumed continuous except the highest one.

We first sketch the Einstein case to establish notation.
Considering a second order discontinuity in the metric across
some characteristic surface $\v=0$, $\d_2 g_{\mn}=\p_{\mn}$,
we have $( \v_{\m} \equiv \partial_{\m} \v )$
\[ \d_1 \G^{\l}~_{\mn} = \frac{1}{2} 
( \v_{\m} \p^{\l}~_{\n} + \v_{\n} \p^{\l}~_{\m}
- \v^{\l} \p_{\mn} ) \;\;, \] 
\begin{eqnarray*}
\d_0 R_{\mn} & = & \v_{\l} (\d_1 \G^{\l}~_{\mn}) -
                   \v_{\n} (\d_1 \G^{\l}~_{\l\m}) \\
             & = & \frac{1}{2} ( \v_{\m} \v_{\l} \p^{\l}~_{\n} +
\v_{\n} \v_{\l} \p^{\l}~_{\m} - \v_{\m} \v_{\n} \p^{\l}~_{\l} -
\v_{\l} \v^{\l} \p_{\mn} )                   
\end{eqnarray*}
and 
\[ \d_0 R = g^{\mn} (\d_0 R_{\mn}) = 
\v^{\m} \v^{\n} \p_{\mn} - \v_{\m} \v^{\m} \p^{\n}~_{\n} \]
which implies for 
\[ \d_0 G_{\mn} = \d_0 (R_{\mn} - \frac{1}{2} g_{\mn} R) =
   \d_0 R_{\mn} - \frac{1}{2} g_{\mn} \d_0 R = 0  \] 
\be 
\d_0 G_{\mn} = \frac{1}{2} \left[ \v_{\m} \v_{\l} \p^{\l}~_{\n} +
\v_{\n} \v_{\l} \p^{\l}~_{\m} - \v_{\m} \v_{\n} \p^{\l}~_{\l} -
\v_{\l} \v^{\l} \p_{\mn} - g_{\mn} 
(\v^{\s} \v^{\t} \p_{\s\t} - \v_{\s} \v^{\s} \p^{\t}~_{\t}) 
\right] = 0 \;\;. \label{ein}
\ee

In the harmonic gauge \( g^{\mn} \G^{\s}~_{\mn} = 0 \), one finds that
its first discontinuity implies
\be
2 \p^{\mn} \v_{\m} - \p^{\m}~_{\m} \v^{\n} = 0  \label{gbir}
\ee
Multiplying this by \( g_{\n\s} \v_{\t} + g_{\n\t} \v_{\s} \),
one gets
\be
\v_{\m} \v_{\l} \p^{\l}~_{\n} + \v_{\n} \v_{\l} \p^{\l}~_{\m} - 
\v_{\m} \v_{\n} \p^{\l}~_{\l} = 0 \;\;,  \label{giki}
\ee
whereas contracting by $\v_{\n}$, one finds
\be
\v^{\m} \v^{\n} \p_{\mn} = \frac{1}{2} \v_{\m} \v^{\m} \p^{\n}~_{\n}
\;\; .  \label{guc} 
\ee
Using (\ref{giki}) and (\ref{guc}) in (\ref{ein}), one ends up with
\[ \d_0 G_{\mn} = \frac{1}{2} ( \v_{\l} \v^{\l} \p^{\mn}
+ \frac{1}{2} g_{\mn} \v_{\l} \v^{\l} \p^{\s}~_{\s} ) = 0 \;\; . \]
Hence taking the trace
\[ \d_0 G^{\m}~_{\m} = \frac{(D+2)}{4} 
\v_{\l} \v^{\l} \p^{\s}~_{\s} = 0 \;\; . \]
The discontinuity in $g_{\mn}$ is arbitrary, hence 
$ \p^{\s}~_{\s} \neq 0 $, which implies that
$\v_{\l} \v^{\l}=0$. This tells that the characteristic surfaces are
null: the discontinuities travel with the speed of light in all
directions. The same holds for the linearized version of the theory
as well of course.

For generic quadratic Lagrangians $ (p R_{\mn} R^{\mn} - q R^2) 
\sqrt{-g}$ in $D=4$, using similar steps (writing the field 
equations, choosing harmonic gauge as before and utilizing the identities 
(\ref{giki}), (\ref{guc})) one finds that $( Q \equiv \v^{\l} \v_{\l} \;,
\p \equiv \p^{\l}~_{\l} )$
\be
Q \left( \frac{1}{2} (p-2q) \v_{\m} \v_{\n} \p -
\frac{p}{2} Q \p_{\mn} - \frac{1}{2} g_{\mn} (\frac{p}{2}-2q) Q \p 
\right) = 0 \;\; . \label{ste}
\ee
Taking the trace, one gets $Q^2 \p (p-3q) = 0$. (The choice $p=3q$
corresponds to Weyl--tensor squared; the scalar degree of freedom
is absent.) For $p=3q$, (\ref{ste}) becomes
\[ q Q ( \frac{1}{2} \v_{\m} \v_{\n} \p - \frac{3}{2} Q \p_{\mn}
+ \frac{1}{4} g_{\mn} Q \p ) = 0 \;\; . \]
Since $\p_{\mn}$ is arbitrary, we see that again $Q=0$, as in Einstein,
so $Q=0$ characterizes both Einstein and the quadratic action.

Finally, we consider the class of actions $\int d^D x f(R) \sqrt{-g}$ 
in $D \geq 4$, whose field equations are
\[ E_{\mn} \equiv R_{\mn} f^\prime - \frac{1}{2} g_{\mn} f +
(g_{\mn} \nabla_{\s} \nabla^{\s} - \nabla_{\m} \nabla_{\n})
f^\prime = 0 \;\; . \]
Hence the order of highest derivatives is four. 
Following similar steps by taking $\d_4 g_{\mn}=\p_{\mn}$, we find 
the same expressions for $\d_3 \G^{\l}~_{\mn}$ and $\d_2 R_{\mn}$ as for 
$\d_1 \G^{\l}~_{\mn}$ and $\d_0 R_{\mn}$ in the Einstein case.
Using these, we get
\[ \d_0 E_{\mn} = (Q g_{\mn} - \v_{\m} \v_{\n})
(\v^{\s} \v^{\t} \p_{\s\t} - Q \p ) f^{\prime\prime} = 0 \;\; . \]
Going to harmonic gauge with identity (\ref{guc}) and
taking the trace, one gets
\[ \d_0 E^{\m}~_{\m} = \frac{(1-D)}{2} Q^2 \p f^{\prime\prime} = 0 \; . \] 
Here too $Q=0$ is the only solution, and so for a wide class of
gravitational actions the propagation obeys the Einstein behavior as well.
As is well known, these systems are variants of Brans-Dicke scalar-tensor 
theories so their ``good propagation" is not surprising.

{\Large{{\bf Acknowledgments}}}

We thank S. Deser for suggesting this problem and collaborating on its applications. \"{O}. S. would also like to thank I. P. Ennes and H. Rhedin for useful discussions. This work was supported in part by NSF, 
under grant no PHY-9315811.

\begin{flushleft}
\Large{{\bf Appendix A}} 
\end{flushleft}

\renewcommand{\theequation}{A.\arabic{equation}}
\setcounter{equation}{0}

In this Appendix, it is shown how the CE requirement (\ref{norm}) causes the
coefficients of discontinuity to evolve according to a linear ODE.

Consider the wavefront at the boundary of a region with smooth enough solution $\oU$.
The following derivation fills a gap in \cite{boi} and generalizes \cite{cv} where
the evolution of discontinuities in first derivatives of the dependent variables
is studied. Choose a root of the characteristic equation, $p_0=h^{I_0}(U,p_i)$. Differentiating (\ref{gen}) with respect to $\v$, and contracting with the corresponding left eigenvector, we have $(A,B,C=1, \ldots ,N)$ 
\be
(\pa_{\v} U_C) \, L^I_A \, (\nabla_{U_C} \cA^{\m}_{AB}) \, (\pa_{\m} U_B) +
L^I_A \, \cA^{\m}_{AB} \, (\pa_{\v} \pa_{\m} U_B) + (\pa_{\v} U_C) \, L^I_A \,
(\nabla_{U_C} \cB_A) = 0 \;\; . \label{is1}
\ee
Now we can take the discontinuity of this equation. We have higher derivative terms,
but notice for the term in the middle that $(\v_{\m} = \pa_{\m} \v)$
\be
\pa_{\v} \pa_{\m} U_B = \v_{\m} (\pa^2_{\v} U_B) + (\pa_{\v} \v_{\m}) (\pa_{\v} U_B) + (\pa_{\m} \psi^i)(\pa_{\psi^i} \pa_{\v} U_B) + (\pa_{\v} \pa_{\m} \psi^i) (\pa_{\psi^i} U_B) \;\; . \label{is2}
\ee
The first term on the right hand side of (\ref{is2}) vanishes when contracted with $\cA^{\m}$ against the left eigenvector. Thus, there is just one $\v$ derivative (i.e. no $\pa^2_{\v}$ pieces), and the discontinuity can be taken as before. We first compute the following to use for the first term in (\ref{is1})
\be
[(\pa_{\v} U_C) \, (\pa_{\m} U_B)] = (\d U_C) \, \v_{\m} \, (\d U_B) + (\d U_C) \, (\pa_{\m} \oU_B) + (\pa_{\v} \oU_C) \, \v_{\m} \, (\d U_B) \;\; . \label{is3}
\ee
Now using (\ref{piR}), (\ref{is3}) and taking the discontinuity of (\ref{is1}), we find
\be
L^I_A \, \cA^{\m}_{AB} \, (\pa_{\m} \psi^i)(\pa_{\psi^i} \d U_B) + m^I_{\,J} \p^J +
\v_{\m} \, (\d U_C) \, L^I_A \, (\nabla_{U_C} \cA^{\m}_{AB}) \, (\d U_B) = 0 \;\; . \label{is4}
\ee
[Here the first term comes from the third term in (\ref{is2}), the last term comes from the first term in (\ref{is3}) and we have collected as ``$m$", the coefficients of terms linear in $\p$ without derivatives. ``$m^I_{\,J}$" are determined by the background solution, as well as the extrinsic geometry of the characteristic surface.] Let's examine the other terms in (\ref{is4}).

The first term in (\ref{is4}) is (up to a redefinition of the coefficient matrix $m$)
\be
(L^I_A \, \cA^{\m}_{AB} \, R^J_B) \, (\pa_{\m} \psi^i)(\pa_{\psi^i} \p_J) = (L^I_A \, \cA^{\m}_{AB} \, R^J_B) \, (\pa_{\m} \p_J) \;\; . \label{is5}
\ee
By taking the $p_i$ derivative (i.e. applying $\pa_{p_i}$) of the straightforward equation $L^I_A \, \cA^{\m}_{AB} \, R^J_B \, p_{\m} = 0$, and using (\ref{ortho}), one finds
\be
L^I_A \, \cA^{i}_{AB} \, R^J_B = - \, \d^{IJ} \frac{\pa h^{I_0}}{\pa p_i} \;\; . \label{is6}
\ee
Hence using the equations for the trajectories (\ref{xps}), the first term in (\ref{is4}) reduces to $\frac{d \p}{d s}$, where
\( \frac{d}{d s} = \frac{\pa}{\pa t} - \frac{d x^i}{d s} \frac{\pa}{\pa x^i} \).

For the last term in (\ref{is4}), we have (by making use of $L^I_A \, \cA^{\m}_{AB} \, R^J_B \, p_{\m} = 0$)
\begin{eqnarray*} 
\v_{\m} \, (\d U_C) \, L^I_A \, (\nabla_{U_C} \cA^{\m}_{AB}) \, (\d U_B) & = &
- \, (\d U_C) \, L^I_A \, \cA^{\m}_{AB} \, (\d U_B) \, (\nabla_{U_C} \v_{\m})
\end{eqnarray*}
Notice that the last factor has $U$ dependence via the characteristic root $p_0$. Hence using (\ref{lamb}) and (\ref{ortho}) (with (\ref{piR}))
\begin{eqnarray}
\v_{\m} \, (\d U_C) \, L^I_A \, (\nabla_{U_C} \cA^{\m}_{AB}) \, (\d U_B) & = & 
(\d U_C) \, \d^{IJ} \, \p_J \, | \vec{\nabla} \v | \, (\nabla_{U_C} \l ) \nonumber \\
& = & | \vec{\nabla} \v | \, \p^I \, \p_J \, R^J_C \, (\nabla_{U_C} \l ) \label{is7}
\end{eqnarray}

Finally then, we have a nonlinear equation for the evolution of the coefficients of discontinuity along rays,
\be
\frac{d \p^I}{d s} + m^I_{\,J} \p^J + | \vec{\nabla} \v | \, \p^I \, \p_J \, R^J_C \, (\nabla_{U_C} \l ) = 0 \;\; . \label{dpis}
\ee
This is computable because all ``$U$'s" above are actually ``$\oU$'s".

Thus, we recognize that CE condition can also be viewed as the statement that the coefficients of discontinuity evolve according to a linear ODE. 

\begin{flushleft}
\Large{{\bf Appendix B}} 
\end{flushleft}

\renewcommand{\theequation}{B.\arabic{equation}}
\setcounter{equation}{0}

In this Appendix, we show the general outline of how (\ref{norm}) is carried out for
models of electrodynamics that depend only on the Maxwell invariant, i.e. $L=L(\a)$. 

By taking ${\bf U} = ({\bf E}, {\bf B})$ and
looking only at the spatial components of the field equation 
\( \pa_{\n} (F^{\m\n} \, L^\prime)=0 \)
and the Bianchi identity $\pa_{\n} \, ^*\!F^{\m\n}=0$ (i.e. setting
$\m=i$), we can write this system in the form
\( {\bf H}^{\m} \; \frac{\pa {\bf U}}{\pa x^{\m}} =0 \)
where ${\bf H}^{\m}$ are $6 \times 6$ matrices. For this new
system (as in the scalar field case when we had 
2$(=4-2\, \cdot \,1)$ nontrivial eigenvalues corresponding 
to the pair of canonical variables for the
only degree of freedom of the theory) we expect to get
$\l=0$ eigenvalue with multiplicity 2$(=6-2\, \cdot \,2)$ for 
each individual ${\bf H}^{i}$ because of the 2 degrees of freedom. 

Just as was done in the scalar field case, we only take 
${\bf H}^1$ to start with. Hence we have 
\( {\bf H}^{0} \; \frac{\pa {\bf U}}{\pa t} + 
{\bf H}^{1} \; \frac{\pa {\bf U}}{\pa x^{1}} = {\bf 0} \) where
\[ {\bf H}^{0} = \left( \begin{array}{cc}
                        {\bf P} & {\bf Q}   \\
                        {\bf 0} & {\bf I}
                 \end{array}  \right) \;\; {\rm and} \;\;
   {\bf H}^{1} = \left( \begin{array}{cc}
                        {\bf S} & {\bf R}   \\
                        {\bf \s} & {\bf 0}
                 \end{array}  \right) \;\;  \]
which have elements (with $i,j=1,2,3$)
\begin{eqnarray*}
p_{i \, j} & = & 2 E_i E_j L^{\prime\prime} - \d_{ij} L^\prime \; , \\
q_{i \, j} & = & - 2 E_i B_j L^{\prime\prime}   \; , \\
s_{i \, j} & = & 2 \e_{1ik} E_j B_k L^{\prime\prime}   \; , \\
r_{i \, j} & = & - \e_{1ik} \left( 2 B_j B_k L^{\prime\prime} 
                            + \d_{jk} L^\prime \right) \; , \\
\s_{i \, j} & = & - \e_{1ik} \; ,
\end{eqnarray*}
and ${\bf I}$ is the $3 \times 3$ identity matrix.

Multiplying by 
\[ ({\bf H}^{0})^{-1} = \left( \begin{array}{cc}
                             {\bf P}^{-1} & -{\bf P}^{-1} \, {\bf Q}  \\
                             {\bf 0} & {\bf I}
                               \end{array}  \right) \;\; ,\]
we bring this system into the canonical form
\( {\bf I} \; \frac{\pa {\bf U}}{\pa t} + 
{\bf W} \; \frac{\pa {\bf U}}{\pa x^{1}} = {\bf 0} \) where
\[ {\bf W} = ({\bf H}^{0})^{-1} \, {\bf H}^{1} =
          \left( \begin{array}{cc}
           {\bf P}^{-1} \,({\bf S} - {\bf Q} {\bf \s})
           & {\bf P}^{-1} \, {\bf R}  \\
           {\bf \s} & {\bf 0}
                 \end{array}  \right) \;\; .\]

Then the characteristic polynomial of ${\bf W}$ turns out to be,
just as predicted, of the form $\l^2 (\l^4 + c_3 \l^3 + c_2 \l^2 +
c_1 \l + c_0) = 0.$ The eigenvectors corresponding
to each $\l_s$ can be taken to be
\[ {\bf e}_s = \left( \begin{array}{c} {\bf a} \\ {\bf b}
                      \end{array} \right) \]
where
\[ \l_1=0: {\bf a}_1 = {\bf 0} \;,
{\bf b}_1 = \left( \begin{array}{c} 1 \\ y_2 \\ y_3
                      \end{array} \right) \;\; {\rm with} \;\;
\left( \begin{array}{c} y_2 \\ y_3 \end{array} \right) =
\frac{-1}{r_{22} r_{33} - r_{23} r_{32}}
\left( \begin{array}{cc} r_{33} & -r_{23} \\ 
                        -r_{32} & r_{22} \end{array} \right) 
\left( \begin{array}{c} r_{21} \\ r_{31} \end{array} \right) \]
\[ \l_2=0: 
{\bf a}_2 = \left( \begin{array}{c} 1 \\ 0 \\ 0 \end{array} \right) \;,
{\bf b}_2 = \left( \begin{array}{c} 0 \\ z_2 \\ z_3
                      \end{array} \right) \;\; {\rm with} \;\;
\left( \begin{array}{c} z_2 \\ z_3 \end{array} \right) =
\frac{-1}{r_{22} r_{33} - r_{23} r_{32}}
\left( \begin{array}{cc} r_{33} & -r_{23} \\ 
                         -r_{32} & r_{22} \end{array} \right) 
\left( \begin{array}{c} s_{21} \\ s_{31} \end{array} \right) \]
and for $\l_s \neq 0 \; (s=3,4,5,6)$ 
\[ {\bf a}_s = \left( \begin{array}{c} 
                       0 \\  \l_{s} ( \rho_{22} - \l_{s} \g_{23}) \\ 
                       \l_{s} ( \rho_{23} + \l_{s} (- \l_{s} + \g_{22}))
                      \end{array} \right) \;\;, \;\;
   {\bf b}_s = \frac{1}{\l_s} {\bf \s} {\bf a}_s =
               \left( \begin{array}{c} 
                       0 \\  -(\rho_{23}+\l_s(-\l_s+\g_{22})) \\
                      (\rho_{22}-\l_s \g_{23})  
                      \end{array} \right) \;\; \]
where
\( \g_{ij} \equiv [{\bf P}^{-1} \,({\bf S} - {\bf Q} {\bf \s})]_{ij} \)
and \( \rho_{ij} \equiv [{\bf P}^{-1} \, {\bf R}]_{ij} .\)
           
Clearly these eigenvectors form a linearly independent set. By
differentiating $\l^4 + c_3 \l^3 + c_2 \l^2 +
c_1 \l + c_0 = 0$, we get
\[ \frac{\pa \l_{p}}{\pa U_{s}} =
- \, \frac{(\l_{p})^3 \, \frac{\pa c_3}{\pa U_{s}} +
           (\l_{p})^2 \, \frac{\pa c_2}{\pa U_{s}} +
            \l_{p} \, \frac{\pa c_1}{\pa U_{s}} +
            \frac{\pa c_0}{\pa U_{s}}}
        {4 \, (\l_{p})^3 + 3 \, c_3 \, (\l_{p})^2 + 2 \, c_2 \, \l_{p} 
+ c_1} \;\;. \]
Substituting this into the CE condition (\ref{norm})
\( \; \sum_{s} \frac{\pa \l_{p}}{\pa U_s} \; e_{p, \, s} = 0 \)
gives a polynomial of order 6 in $\l$, but by using 
$\l^4 + c_3 \l^3 + c_2 \l^2 + c_1 \l + c_0 = 0$ repeatedly, 
one can reduce this to a polynomial of order 3, whose 
coefficients must be set equal to zero simultaneously.

Doing so, we find that the only nontrivial covariant
condition, we can impose such that these coefficients
vanish simultaneously, is
\be
L^{\prime} L^{\prime\prime\prime} - 3 (L^{\prime\prime})^2 = 0  \;\; . \label{lofal}
\ee

\begin{flushleft}
\Large{{\bf Appendix C}} 
\end{flushleft}

\renewcommand{\theequation}{D.\arabic{equation}}
\setcounter{equation}{0}

Here, for completeness, we present (\ref{tar3}) and (\ref{tar4}) in terms of $L$ and its derivatives only. They become $(K \equiv \laa \lbb -\lab^2)$
\begin{eqnarray}
\frac{3}{2} \la \labb \left( \la  (16 \laa^3 \lab + 8 \laa \lab^3 + \lab^3 \lbb) \right. \hspace{4cm}  \nonumber \\
\hspace{4cm} \left. - K \left[ 8 \a \laa^2 \lab + \b ( 8 \laa \lab^2 + 4 \laa^2 \lbb + \lab^2 \lbb) \right] \right) \nonumber \\
+ \frac{1}{2} \la \laaa \left( \la \lab  (16 \laa \lab^2 + 8 \lab^2 \lbb + \lbb^3) \right. \hspace{4cm}  \nonumber \\
\hspace{4cm} \left. - K \left[ 8 \a \lab^3 + \b \lbb ( 12 \lab^2 + \lbb^2) \right] \right) \nonumber \\
- \frac{3}{2} \la \lab \laab \left( \la \lab (16 \laa^2 + 4 \lab^2 + 4 \laa \lbb + \lbb^2) \right. \hspace{4cm} \nonumber \\
\hspace{5cm} \left. - K \left[ 8 \a \laa \lab + \b (4 \lab^2 + 8 \laa \lbb + \lbb^2) \right] \right) \nonumber \\
- \frac{1}{2} \la \lbbb \left( \la (16 \laa^4 + 12 \laa^2 \lab^2 + \lab^4 - 4 \laa^3 \lbb) \right. \hspace{4cm} \nonumber \\
\hspace{5cm} \left. - K \left[ 8 \a \laa^3 + \b \lab (12 \laa^2 + \lab^2) \right] \right) \nonumber \\
- \frac{3}{2} (4 \laa + \lbb) K^2 [\la \lab - \b K]  & = & 0 \hspace{.5cm} \label{am1}
\end{eqnarray} 
and
\begin{eqnarray}
- \frac{3}{2} \la \laab \left( 
(4 \laa + \lbb) (2 \la^2 \lab^2 - \a \la \lab^2 \lbb)
\right. \hspace{4cm}  \nonumber \\
\hspace{3cm}+ \b \la \lab 
(16 \laa \lab^2 + 6 \lab^2 \lbb - 2 \laa \lbb^2) \nonumber \\
\hspace{3cm} \left. - \b K \left[ - \a \lab \lbb^2 + 2 \b 
(4 \laa \lab^2 + 2 \lab^2 \lbb + \laa \lbb^2) \right] \right) 
\nonumber \\
+ \frac{3}{2} \la \labb \left( 
(4 \laa^2 + \lab^2) (2 \la^2 \lab - \a \la \lab \lbb)
\right. \hspace{4cm} \nonumber \\
\hspace{3cm} - \b K \lab \left[ -\a \lab \lbb + 2 \b (4 \laa^2 + \lab^2 + 2 \laa \lbb)  \right] \nonumber \\
\hspace{3cm} \left. + 2 \b \la  (8 \laa^2 \lab^2 + 2 \lab^4 + \laa \lbb \lab^2 - \laa^2 \lbb^2) \right) \nonumber \\
+ \frac{1}{2} \la \laaa \left( 
(4 \lab^2 + \lbb^2) (2 \la^2 \lab - \a \la \lab \lbb)
\right. \hspace{4cm} \nonumber \\
\hspace{3cm} 
+ \b \la (16 \lab^4 + 6 \lab^2 \lbb^2 - 2 \laa \lbb^3) \nonumber \\
\hspace{3cm} \left. - \b K (8 \b \lab^3 + 6 \b \lab \lbb^2 - \a \lbb^3)
\right) \nonumber \\
- \frac{1}{2} \la \lbbb \left( 2 \la^2 \laa (4 \laa^2 + 2 \lab^2 - \laa \lbb)
\right. \hspace{4cm} \nonumber \\
\hspace{3cm} + 2 \b \la \laa \lbb (8 \laa^2 + 5 \lab^2 - 3 \laa \lbb)
\nonumber \\
\hspace{3cm} \left. - \b K (8 \b \laa^3 + 6 \b \laa \lab^2 - \a \lab^3) 
- \a \la (\lab^4 + 4 \laa^3 \lbb) 
\right) \nonumber \\
- \frac{3}{2} K^2 (4 \la + 4 \b \lab - \a \lbb) (\la \lab - \b K) & = & 0 
\label{am2}
\end{eqnarray}
respectively.

\end{document}